\documentclass[aps,preprint,epsf,onecolumn]{revtex4}
\usepackage{amsmath}
\usepackage{epsfig}
\usepackage{amssymb}

\begin{document}

\title{Light scattering by a metallic nanoparticle coated with a nematic liquid crystal}

\author{Vassilios~Yannopapas, Nikolaos~G.~Fytas, Vicky~Kyrimi,
  Efthymios~Kallos, Alexandros~G.~Vanakaras, and Demetri~J.~Photinos}

\affiliation{ Department of Materials Science, University of Patras, GR-26504 Patras, Greece}

\date{\today}

\begin{abstract}
We study the optical properties of gold nanoparticles coated with
a nematic liquid crystal whose director field is distributed
around the nanoparticle according to the anchoring conditions at
the surface of the nanoparticle. The distribution of the nematic
liquid crystal is obtained by minimization of the corresponding
Frank free-energy functional whilst the optical response is
calculated by the discrete-dipole approximation.We find, in
particular, that the anisotropy of the nematic liquid-crystal
coating does not affect much the (isotropic) optical response of
the nanoparticle. However, for strong anchoring of the nematic
liquid-crystal molecules on the surface of nanoparticle, the
inhomogeneity of the coating which is manifested by a ring-type
singularity (disclination or Saturn ring), produces an enhancement
of the extinction cross spectrum over the entire visible
spectrum.
\end{abstract}

\maketitle

\section{Introduction}
\label{intro}

Metamaterials are man-made structures with electromagnetic (EM)
properties which are not met in naturally occurring materials,
such as artificial magnetism, negative refractive index (NRI),
near-field amplification, cloaking and other optical illusions.
The holy grail of research in this discipline has been the
realization of metamaterials with the above EM properties in the
visible regime. To this end, top-down technologies and
lithographic techniques have been employed for the realization of
optical metamaterials. However, due to fabrication restrictions in
the top-down approaches, only two dimensional (2D), planar
metamaterials have been realized in the optical regime
\cite{soukoulis}. On the contrary, bottom-up approaches based on
self-assembly technology allow for the realization of true,
three-dimensional (3D) optical metamaterials with the promise of
lower cost, high throughput, and small sensitivity to damage or
fabrication errors.

The most promising way of realizing metamaterials by a bottom-up
approach relies on organic chemistry of mesogens and
macromolecules. Namely, macromolecules such as dendrimers and
mesogens of controlled shape are ideally suited for organizing
metal nanoparticles (NPs) into composite materials with tailored
properties such as particle size, shape, particle distance, etc.
Macromolecular and low-molar liquid crystal (LC) ligands are
attached to the particles by chemical methods and arrangements
such as disks or rods can be envisaged where the NP is located at
the center and is surrounded by radially or tangentially arranged
organic molecules \cite{cseh,zeng,khatua,gardner,kanie,georg}.

In this work, we study the optical response of a single metallic
NP decorated by a nematic liquid crystal (NLC), in order to assess
the effect of the inherent inhomogeneity and anisotropy of the NLC
on the plasmonic excitation of the metallic NP. Our study will be
based on treating the NLC surrounding the NP as a continuous
medium but with the aforementioned inhomogeneity and anisotropy of
the NLC. The anisotropy of the NLC stems from the axial symmetry
of NLC molecules whilst the inhomogeneity is a result of the
anchoring of the NLC molecules on the surface of the NP. The
distribution of the director field of the NLC around a certain NP
is taken by minimization of the corresponding functional of the
elastic Frank free energy around and at the surface of the NP. For
two limiting cases for the anchoring energy, i.e., $W \rightarrow
0$ (weak anchoring) and $W \rightarrow \infty$ (strong anchoring)
we use analytic expressions for the spatial distribution of the
director field. A spherical metallic NP is assumed to be coated by
a spherical layer of a NLC with given director-field distribution.
The EM response of the NLC-coated NP is studied by means of the
discrete-dipole approximation \cite{purcell,flatau,yurkin} using
as input the polarizability tensor at each point in space which is
provided by the distribution of the director field for the NLC
coating (the corresponding polarizability tensor for the metallic
NP/ core is diagonal and constant in space). It is worth noting
that the optical response of NLC-coated metallic NP in free space
\cite{park_1} and  atop a substrate \cite{park_2} has been studied
in the past within, however, a limited frequency range around the
NP surface plasmon (SP) frequency.

\section{Theory}
\label{theory}

\subsection{Distribution of the director field}
\label{director}

In the absence of any material boundaries, a NLC is uniformly
distributed in space. However, when boundaries are present the
director field of the NLC is inhomogeneously distributed in space
according to the anchoring conditions on the physical boundary. In
our case, the physical boundary is the surface of the NP. In order
to find the distribution of the director field ${\bf n}({\bf r})$
around the NP, one minimizes the functional of the Frank free
energy which, in our case, is taken in the one-constant
approximation, i.e.,
\begin{equation}
F=\int \frac{1}{2} K [(\nabla \cdot {\bf n})^2 + (\nabla \times
{\bf n})^{2}] d^{3} r. \label{eq:frank}
\end{equation}
At the surface of NP, the functional of the free energy is taken
in the Rapini \cite{rapini} approximation
\begin{equation}
F_{s}=-\oint \frac{1}{2} W ({\bf n} \cdot \hat{{\bf s}}) dS
\label{eq:rapini}
\end{equation}
where $\hat{{\bf s}}$ is the unit vector normal to the surface and
$W$ and anchoring energy. Eq.~(\ref{eq:rapini}) constitutes the
boundary condition in our case. The functionals of
Eqs.~(\ref{eq:frank}) and (\ref{eq:rapini}) are minimized under
the constraint $| {\bf n}({\bf r})|^{2}=1$ for the director vector
\begin{equation}
{\bf n}({\bf r})= (n_{x},n_{y},n_{z}) = (\sin \beta({\bf r}) \cos
\phi, \sin \beta({\bf r}) \sin \phi, \cos \beta ({\bf r}))
\label{eq:d_field}
\end{equation}
which, obviously, respects the cylindrical symmetry of the problem
($\theta$, $\phi$ are the polar and azimuthal angles of the
spherical coordinate system). $\beta ({\bf r})$ is the angle
between the director vector and the $z$ axis which is taken to be
the optical axis of the NLC in the absence of the NP.

For the case of {\it weak} anchoring, i.e., $W S / K \ll 1$ where
$S$ is the NP radius, minimization of the functionals
(\ref{eq:frank}) and (\ref{eq:rapini}) leads to an analytical
function for $\beta({\bf r})$ \cite{kuksenok}
\begin{equation}
\beta ({\bf r}) = \frac{ W S} {4 K} (\frac{S}{r})^{3} \sin
2\theta. \label{eq:weak}
\end{equation}
The above formula is valid for anchoring energies $W \leq
10^{-4}{\rm  ergs/cm^{2}}$. For a typical value of $K\sim 10^{-6}
{\rm  ergs/cm}$, Eq.~(\ref{eq:weak}) is valid for NP radii $S < 5
\mu$m. \cite{kuksenok}

For the case of {\it strong} anchoring, i.e., $W \rightarrow
\infty$, one can also have an analytic function for $\beta ({\bf
r})$ \cite{kuksenok}
\begin{equation}
\beta({\bf r}) = \theta -\frac{1}{2} \arctan (\frac{\sin 2
\theta}{1/f(r) + \cos 2 \theta}) \label{eq:strong}
\end{equation}
where
\begin{equation}
f(r)=(\frac{r}{a})^{3}+A+B+C \exp(-r/a) \label{eq:anchor_func}
\end{equation}
and
\begin{eqnarray}
A&=&\frac{S^{3}}{a^2 (a-S)^{2}} \biggl[ S-a +(\frac{a}{S})^2
(4a-3S)   \biggl. \nonumber \\ \biggr. & \times &
[\frac{a}{S} \exp (-S/a) - \exp(-1)] \biggl] \nonumber \\
B&=&\frac{S^{3}}{a^2 (a-S)^{2}} \bigl[a-S + (4a-3s) \bigr.
\nonumber \\ \bigl. & \times &
 [\exp (-1) -
\exp(-S/a)] \bigr], \nonumber \\
C&=&-\frac{4a-3S}{a-S}. \label{eq:coeffs}
\end{eqnarray}
$r=a$ is the distance from the center of the NP of a disclination
(Saturn) ring of $(-1/2)$ singularity at the equatorial plane
($\theta=\pi/2$). \cite{kuksenok,terentjev,lubensky} The Saturn
ring is a result of the pinning of the NLC to the surface of the
NP. We note that the above analytical functions for the NLC
director field are valid for two limiting cases (weak and strong
anchoring). For intermediate anchoring energies as well as for
more general Frank free-energy functionals one has to resort to
numerical methods
\cite{park_1,park_2,ruhwandl,stark,grollau,fukuda,cheung}.

\subsection{Discrete dipole approximation}
\label{dda}

We study the optical response of NLC-coated metallic NP using the
discrete dipole approximation (DDA). \cite{purcell,flatau,yurkin}
The NLC-coated NP (scattering object in our case) is considered as
an array of point dipoles ($i=1,\cdots,N$) each of which is
located at the position ${\bf r}_{i}$ and corresponds to a dipole
moment ${\bf P}_{i}$ and a (position-dependent) polarizability
tensor $\tilde{{\bf \alpha}}_{i}$. The above quantities are
connected by
\begin{equation}
{\bf P}_{i}={\tilde {\alpha}_{i}} {\bf E}_{i} \label{eq:pae}
\end{equation}
where ${\bf E}_{i}$ is the electric field at $i$-th dipole,
\begin{equation}
{\bf E}_{i}={\bf E}_{inc,i} - \sum_{j \neq i}{\bf A}_{ij} \cdot
{\bf P}_{j} \label{eq:local_field}
\end{equation}
which is the sum of the directly incident field ${\bf E}_{inc,i}$
as well as the field scattered by all the other dipoles $j \neq i$
and it is incident on the $i$-th dipole [second term of
Eq.~(\ref{eq:local_field})]. The interaction matrix ${\bf A}_{ij}$
is given from
\begin{equation}
{\bf A}_{ij} = \frac{\exp(ik r_{ij})}{r_{ij}} \biggl[k^{2}
(\hat{r}_{ij} \hat{r}_{ij} - {\bf 1}_{3}) + \frac{i k r_{ij}
-1}{r_{ij}^{2}} ( 3\hat{r}_{ij} \hat{r}_{ij} - {\bf 1}_{3})
\biggr], \ \ i \neq j \label{eq:interaction}
\end{equation}
where ${\bf 1}_{3}$ is the $3 \times 3$ unit matrix, ${\bf
r}_{ij}={\bf r}_{i} - {\bf r}_{j}$, $\hat{{\bf r}}_{ij}={\bf
r}_{ij}/ |{\bf r}_{ij}|$. By combining Eqs.~(\ref{eq:pae}) to
(\ref{eq:interaction}) we obtain a linear system of equations,
i.e.,
\begin{equation}
\sum_{j=1}^{N} {\bf A}_{ij} {\bf P}_{j} = {\bf E}_{inc,i}
\label{eq:dda_system}
\end{equation}
where the diagonal elements of the interaction matrix are
essentially the inverse of the polarizability tensor of each
dipole, i.e.,
\begin{equation}
{\bf A}_{ii}=[\tilde{\alpha}_{i}]^{-1}.
\label{eq:diag_interaction}
\end{equation}
For an anisotropic sphere characterized by a dielectric tensor
$\tilde{\epsilon}_{s}$ and is immersed within an isotropic host of
dielectric constant $\epsilon_{h}$, the polarizability tensor of
the sphere is given by the Clausius-Mossoti formula for
anisotropic spheres \cite{levy}, i.e.,
\begin{equation}
\tilde{\alpha}_{i}=V_{s} \frac{3 \epsilon_{h}} {4 \pi}
[\tilde{\epsilon}_{s}-\epsilon_{h} {\bf
1}_{3}][\tilde{\epsilon}_{s}+2 \epsilon_{h} {\bf 1}_{3}]^{-1}.
\label{eq:clausius_mossoti}
\end{equation}
For a nematic liquid crystal $\tilde{\epsilon}_{s}=
diag(\epsilon_{\parallel},\epsilon_{\parallel},\epsilon_{\perp})$
while for a metallic sphere $\tilde{\epsilon}_{s}=\epsilon_{m}
{\bf 1}_{3}$. Eq.~(\ref{eq:clausius_mossoti}) provides the
polarizability of a NLC when the director vector is oriented along
the $z$-axis of the (global) coordinate system. When the director
vector of the $i$-th dipole forms arbitrary angles $\alpha$,
$\beta$ and ${\gamma}$ with the $x$, $y$, and $z$ axes,
respectively, the polarizability tensor is given by \cite{loiko}
\begin{equation}
\hat{\alpha}= {\bf M}^{-1} \tilde{\alpha} {\bf M}
\label{eq:rotations}
\end{equation}
where
\begin{equation}
{\bf M}= {\bf R}_{x}(\alpha) {\bf R}_{y}(\beta) {\bf
R}_{z}(\gamma) \label{rot_matr}
\end{equation}
and ${\bf R}_{x}(\alpha)$, ${\bf R}_{y}(\beta)$, ${\bf
R}_{z}(\gamma)$ are the rotation matrices about the $x$, $y$, and
$z$ axes, respectively.

Having determined the dipole moment ${\bf P}_{i}$ at each point
dipole, one can calculate quantities such as the scattering,
extinction and absorption cross sections, i.e.,
\begin{eqnarray}
C_{sc}&=& \frac{4 \pi k}{|E_{inc}|^{2}} \sum_{i=1}^{N}
\biggl[\frac{2}{3} k^{2} |{\bf P}_{i}|^{2} - \Im ({\bf P}_{i}
\cdot {\bf
E}_{self,i}^{*}) \biggr] \\
C_{ext}&=&4 \pi k \sum_{i=1}^{N} \Im ({\bf P}_{i} \cdot {\bf
E}_{inc,i}^{*}) \\
C_{abs}&=&4 \pi k \sum_{i=1}^{N} \Im ({\bf P}_{i} \cdot {\bf
E}_{i}^{*}) \label{eq:cross_secs}
\end{eqnarray}
where ${\bf E}_{self,i} = {\bf E}_{i} - {\bf E}_{inc,i}$.

\begin{figure}[h]
\centerline{\includegraphics[width=5cm]{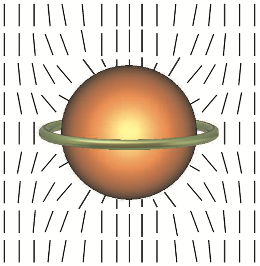}}
\caption{A metallic nanoparticle coated with a strongly anchored
NLC and its disclination ring at the equatorial plane.}
\label{fig1}
\end{figure}

\subsection{Nonlocal optical response for the gold nanoparticle}

Due to the small size (below 10~nm) of the fabricated NLC-coated
gold NPs, the dielectric function of gold appearing in
Eq.~(\ref{eq:clausius_mossoti}) should be provided within a
nonlocal description of the optical response of such NPs
\cite{ruppin,agarwal,claro,liebsch,abajo,chen}. At the same time,
a realistic description of the optical properties of NLC-coated
NPs requires the adoption of an experimentally obtained dielectric
function \cite{jk} which, apart from the Drude-type behavior, it
includes the contribution of interband transitions. Lastly, due to
the small size of the NP, scattering of the free electrons at the
boundaries of the NP should be considered as well. Taking all the
above into account, the dielectric function of gold is written
\cite{abajo},
\begin{equation}
\epsilon_{m}(\omega,k)=\epsilon_{exp}(\omega)-\epsilon_{Drude}(\omega)+\epsilon_{NL}(\omega,k)
\label{eq:eps_metal}
\end{equation}
where $\epsilon_{exp}$ is the experimental dielectric function of
gold \cite{jk} and
\begin{equation}
\epsilon_{Drude} = 1 - \frac{\omega_{p}^{2}}{\omega (\omega + i
\gamma_{bulk})} \label{eq:drude}
\end{equation}
is the Drude-type dielectric function where $\omega_{p}=8.99$~eV
and $\gamma_{bulk}=0.05$~eV are the plasma frequency and the loss
factor of bulk gold, respectively. The nonlocal part of the
dielectric function of Eq.~(\ref{eq:eps_metal}) is provided by the
hydrodynamic model \cite{ruppin,claro,chen}
\begin{equation}
\epsilon_{NL} = 1 - \frac{\omega_{p}^{2}}{\omega^2 + i \omega
\gamma_{NP} - \beta k^{2}} \label{eq:drude_NL}
\end{equation}
where $\beta = (3/5) v_{F}^{2}$  with $v_{F}=0.903~{\rm eV}\cdot
{\rm nm}$ being the electron Fermi velocity for gold.
$\gamma_{NP}$ is the loss factor (inverse relaxation time of the
free electrons) corrected in order to take into account the
scattering of the electrons at the spherical boundary of the NP
\cite{grady}
\begin{equation}
\gamma_{NP}=\gamma_{bulk} + G v_{F} / S \label{eq:corr_loss}
\end{equation}
where $S$ is the NP radius and $G$ is a dimensionless fitting
parameter which describes the nature of the electron-surface
scattering \cite{grady}. In Figs.~\ref{fig2}-\ref{fig4} we have
taken $G=1$ whilst in Fig.~\ref{fig5} we have taken $G=3$ for best
fitting to the experimental data.

Once a nonlocal dielectric function is assumed for the gold NP,
the corresponding polarizability must be given in the same
(nonlocal) framework, i.e., $\tilde{\alpha}_{i}$ of
Eq.~(\ref{eq:clausius_mossoti}) is provided by
\cite{claro,abajo,chen}
\begin{equation}
\tilde{\alpha}_{i}=V_{s} \frac{3 \epsilon_{h}} {4 \pi}
[\tilde{\zeta}_{s}-\epsilon_{h} {\bf 1}_{3}][\tilde{\zeta}_{s}+2
\epsilon_{h} {\bf 1}_{3}]^{-1}. \label{eq:clausius_mossoti_NL}
\end{equation}
with
\begin{equation}
\tilde{\zeta}_{s}=\Bigl[ \frac{6S}{\pi} \int_{0}^{\infty} dk
\frac{j_{1}^{2} (k S)}{\epsilon_{m}(\omega,k)}\Bigr]^{-1}
\label{eq:polariz_NL}
\end{equation}
where $j_{1}$ is the spherical Bessel function for $\ell=1$ and
$\epsilon_{m}$ is given by Eq.~(\ref{eq:eps_metal}). Thanks to the
homogeneous and isotropic nature of the NPs studied here as well
as their small size ($<10$~nm), a gold NP can be very well
approximated as a {\it single} point dipole in
Eqs.~(\ref{eq:dda_system}) whose polarizability is provided by the
nonlocal polarizability of Eq.~(\ref{eq:polariz_NL}). This speeds
up the convergence in the DDA calculations [numerical solution of
Eqs.~(\ref{eq:dda_system})] since the gold nanoparticle need not
be further discretized to smaller dipoles.

\section{Results and discussion}
\label{results}

\begin{figure}[h]
\centerline{\includegraphics[width=7cm]{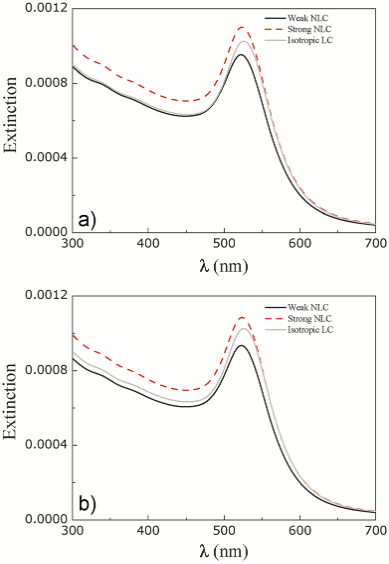}}
\caption{Extinction spectrum for a 7.5~nm gold nanoparticle coated
with E7 NLC of 5~nm thickness for weak (solid line) and strong
(broken line) anchoring conditions. The light grey line
corresponds to the case of an isotropic medium with
$n_{iso}=(2n_{\parallel} + n_{\perp})/3$. For (a) light is
incident along the optical axis of the NLC (taken to be the
$z$-axis) while for (b) light is incident normal to the optical
axis).} \label{fig2}
\label{fig1}
\end{figure}

We consider gold NPs coated by E7 commercially available NLC with
refractive indices $n_{\parallel}=1.74$ and $n_{\perp}=1.53$
\cite{li}. Fig.~\ref{fig1} shows a gold NP surrounded by a Saturn
ring of $-1/2$ singularity which is present under strong anchoring
conditions. Fig.~\ref{fig2} shows the extinction spectrum for gold
NP of 7.5~nm radius surrounded by a NLC coating of 5~nm thickness
for right-circularly polarized light incident along
(Fig.~\ref{fig2}a) and normal (Fig.~\ref{fig2}b) to the NLC
optical axis.

We note that the scattering spectrum is much smaller than the
absorption spectrum, i.e., extinction almost exclusively stems
from absorption.  The calculated extinction spectra are for weak
($W S / K = 2 \times 10^{-4}$) and strong anchoring conditions as
well as for the case of an isotropic medium whose refractive index
is the average of two refractive indices of the NLC, i.e.,
$n_{iso}=(2n_{\parallel} + n_{\perp})/3$.

\begin{figure}[h]
\centerline{\includegraphics[width=7cm]{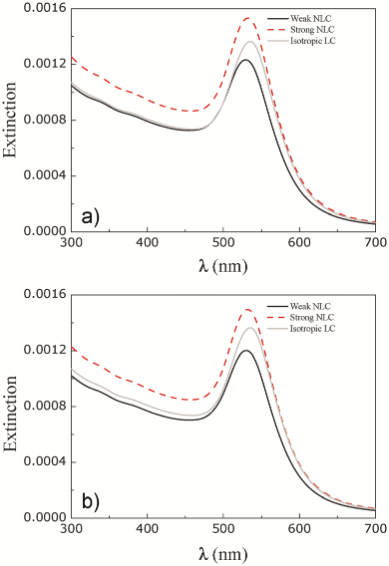}}
\caption{The same as Fig.~\ref{fig2} but for a NLC coating of
7.5~nm thickness.} \label{fig3}
\label{fig1}
\end{figure}

\begin{figure}[h]
\centerline{\vbox{\includegraphics[width=7cm]{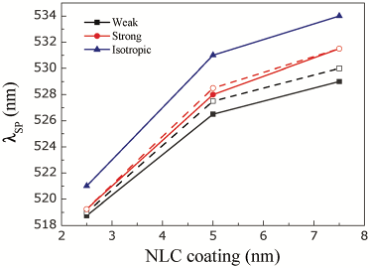}}}
\caption{Position of the SP peak for different configurations of
the NLC director field, for light incident along (solid lines -
filled symbols) and normal (broken lines - empty symbols) to the
optical axis of the NLC.} \label{fig4}
\end{figure}

Fig.~\ref{fig3} shows extinction spectra similar to those of
\ref{fig2} but for a thicker coating, i.e., 7.5~nm thickness. The
peak in all extinction spectra stems from the excitation of the SP
resonance of the gold NP. Evidently, for strong anchoring
conditions, the extinction spectrum is on the average 20\% higher
than the two other types of coating (weakly anchored NLC and
isotropic). Fig.~\ref{fig4} shows the position of the SP peak as a
function of the coating thickness, for different configurations of
the NLC director field and different directions of light
incidence. As the coating becomes thicker, the difference in the
SP position between the isotropic LC and the NLC phases increases,
i.e., for a 7.5~nm thickness and for incidence along the NLC axis,
the SP peak for weak anchoring conditions is 5~nm below the SP
peak corresponding to the isotropic LC coating.

It can also be seen that the position of the SP depends on the
incidence direction (and hence on polarization) as a result of the
inherent anisotropy of the NLC coating which induces the splitting
of the SP mode \cite{park_1,park_2}. The splitting is somewhat
larger for the weak anchoring conditions implying a more
anisotropic profile for the director-field distribution. As
expected, the SP splitting is zero for the isotropic LC phase.

The small differences in the position (about 1~nm) and magnitude
(about 2\%) of the extinction spectra for both cases of incidence
(parallel and normal to the optical axis of the NLC) indicates
that the anisotropy of the NLC does not affect much the
(isotropic) EM response of the NP. This means that in a periodic
metamaterial structure made from NLC-functionalized metallic NPs,
if anisotropic optical response is experimentally measured it will
be mostly attributed to the particular arrangement of the NPs in
space and very little to the anisotropy of the surrounding medium
(NLC) which will only have the role of a 'glue' which keeps the
NPs joint together in a certain geometrical structure. Of course,
as mentioned above, the inhomogeneous distribution of the NLC
around the metallic NP should be taken into account as it results
in an overall 20\% higher extinction spectrum for strong anchoring
conditions and a considerable SP shift of about 5~nm for weak
anchoring conditions. We note, however, than neither the
anisotropy nor the inhomogeneity of the NLC coating produces new
structure, e.g., additional peaks, in the extinction spectrum as a
result of the small anisotropy of the refractive index.

\begin{figure}[h]
\centerline{\vbox{\includegraphics[width=7cm]{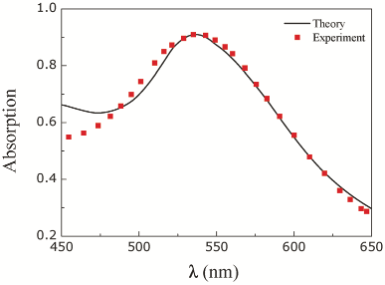}}}
\caption{Theoretical (line) and experimental (squares) absorption
spectra for a 6~nm gold NP within the NLC
4-cyano-4-n-pentylbiphenyl (5CB). The theoretical curve is taken
by assuming $G=3$ in Eq.~(\ref{eq:corr_loss})} \label{fig5}
\end{figure}

Finally, in Fig.~\ref{fig5} we show the absorption spectrum for a
gold NP in the NLC 4-cyano-4-n-pentylbiphenyl (5CB) along with
corresponding experimental data from Ref.~\cite{khatua}. The
agreement between theory and experiment is very good except for
wavelengths shorter than 475~nm  where the theoretical curve is an
decreasing function of wavelength while the experimental one is
increasing. It is worth noting that our theoretical model cannot
capture experimental conditions such as the free rotation and the
size distribution of the NPs.

\section{Conclusions}
We have studied the optical response of gold nanoparticles coated
with a nematic liquid crystal. The distribution of the director
field of the nematic liquid crystal was taken from analytic
functions determined by minimization of the elastic Frank free
energy. The optical response has been calculated by the discrete
dipole approximation which takes into account both the anisotropy
and inhomogeneity of the liquid-crystal coating. We have found
that the anisotropy of the coating has small influence on the
(isotropic) response of the nanoparticle. When the liquid-crystal
molecules are strongly anchored at the surface of the
nanoparticle, the distribution of the nematic liquid crystal
possess a topological singularity (Saturn ring) and has a
significant impact on the magnitude of the extinction spectrum of
the nanoparticle. For weakly anchored liquid-crystal molecules,
the surface-plasmon peak is shifted a few nm relative to a coating
of isotropic liquid-crystal phase.

\acknowledgments{This work has been supported by the European Community's Seventh
Framework Programme (FP7/2007-2013) under Grant Agreement No.
228455-NANOGOLD (Self-organized nanomaterials for tailored optical
and electrical properties).}


\begin{thebibliography}{999}
\bibitem{soukoulis} C.~M.~Soukoulis and M.~Wegener, Nat.~Phot.
{\bf 5}, 523 (2011).
\bibitem{cseh} L.~Cseh and G.~H.~Mehl, J.~Am.~Chem.~Soc. {\bf 128}, 13376 (2006).
\bibitem{zeng} X.~Zeng, F.~Liu, A.~G.~Fowler, G.~Ungar,
L.~Cseh, G.~H.~Mehl, and J.~E.~Macdonald, Adv.~Mater. {\bf 21},
1746 (2009).
\bibitem{khatua} S.~Khatua, P.~Manna, W.~-S.~Chang, A.~Tcherniak,
E.~Friedlander, E.~R.~Zubarev, and S.~Link, J.~Phys.~Chem.~C {\bf
114}, 7251 (2010)
\bibitem{gardner} Q.~Liu, Y.~Cui, D.~Gardner, X.~Li, S.~He, and
I.~I.~Smalyukh, Nano~Lett. {\bf 10}, 1346 (2010).
\bibitem{kanie} K.~Kanie, M.~Matsubara, X.~Zeng, F.~Liu,
G.~Ungar, H.~Nakamura, and A.~Muramatsu, J.~Am.~Chem.~Soc.  {\bf
134}, 808 (2011).
\bibitem{georg} C.~H.~Yu, C.~P.~J.~Schubert, C.~Welch, B.~J.~Tang,
M.~-G.~Tamba, and G.~H.~Mehl, J.~Am.~Chem.~Soc. {\bf 134}, 5076
(2012).
\bibitem{purcell} E.~M.~Purcell and C.~R.~Pennypacker,
Astrophys.~J. {\bf 186}, 705 (1973).
\bibitem{flatau} P.~J.~Flatau, Opt.~Lett. {\bf 22},
1205 (1997).
\bibitem{yurkin} M.~A.~Yurkin, and A.~G.~Hoekstra,
J.~Quant.~Spec.~Rad.~Transfer {\bf 106}, 558 (2007).
\bibitem{park_1} S.~Y.~Park and D.~Stroud, Appl.~Phys.~Lett. {\bf
85}, 2920 (2004).
\bibitem{park_2} S.~Y.~Park and D.~Stroud, Phys.~Rev.~Lett. {\bf 94},
217401 (2005).
\bibitem{rapini} A.~Rapini and M.~Popoular, J.~Phys.~(Paris)
~Colloq. {\bf 30}, C-4-54 (1969).
\bibitem{kuksenok} O.~V.~Kuksenok, R.~W.~Ruhwandl,
S.~V.~Shiyanovskii, E.~M.~Terentjev, Phys.~Rev.~E {\bf 54}, 5198
(1996).
\bibitem{terentjev} E.~M.~Terentjev, Phys.~Rev.~E {\bf 51}, 1330
(1995).
\bibitem{lubensky} T.~C.~Lubensky, D.~Pettey, N.~Currier,
and H.~Stark, Phys.~Rev.~E {\bf 57}, 610 (1998).
\bibitem{ruhwandl} R.~W.~Ruhwandl and E.~M.~Terentjev, Phys.~Rev.~E
{\bf 56}, 5561 (1997).
\bibitem{stark} H.~Stark, Eur.~Phys.~J.~B {\bf 10}, 311 (1999).
\bibitem{grollau} S.~Grollau, N.~L.~Abbott, and J.~J.~de~Pablo,
Phys.~Rev.~E {\bf 67}, 011702 (2003).
\bibitem{fukuda} J.~Fukuda, M.~Yoneya, and H.~Yokoyama, Eur.~Phys.~J.~E
{\bf 13}, 87 (2004).
\bibitem{cheung} D.~L.~Cheung and M.~P.~Allen, Phys.~Rev.~E {\bf 74},
021701(2006).
\bibitem{levy} O.~Levy and D.~Stroud, Phys.~Rev.~B {\bf
56}, 8035 (1997).
\bibitem{loiko}  V.~A.~Loiko, V.~I.~Molochko, Appl.~Opt. {\bf 38},
2857 (1999).
\bibitem{ruppin} R.~Ruppin, Phys.~Rev.~B {\bf 11}, 2871 (1975).
\bibitem{agarwal} G.~S.~Agarwal and S.~V.~O'~Neil, Phys.~Rev.~B
{\bf 28}, 487 (1983).
\bibitem{claro} R.~Fuchs and F.~Claro, Phys.~Rev.~B {\bf 35}, 3722
(1987).
\bibitem{liebsch} A.~Liebsch, Phys.~Rev.~B {\bf 48}, 11317 (1993).
\bibitem{abajo} F.~J.~Garc\'{i}a~de~Abajo, J.~Phys.~Chem.~C
{\bf 112}, 17983 (2008).
\bibitem{chen} C.~W.~Chen, L.~S.~Liao, H.~-~P.~Chiang, and
P.~T.~Leung, Appl.~Phys.~B {\bf 99}, 223 (2010).
\bibitem{jk} R.~B.~Johnson and R.~W.~Christy, Phys.~Rev.~B {\bf
6}, 4370 (1972).
\bibitem{grady} N.~K.~Grady, N.~J.~Halas, and P.~Nordlander,
Chem.~Phys.~Lett. {\bf 399}, 167 (2004).
\bibitem{li} J.~Li, Y.~Ma~Y, Y.~Gu, I.~C.~Khoo, and Q.~Gong,
Appl.~Phys.~Lett., {\bf 98}, 213101 (2011).
\end{thebibliography}
\end{document}